\documentstyle[11pt,newpasp,twoside,psfig]{article}
\def\edcomment#1{\iffalse\marginpar{\raggedright\sl#1\/}\else\relax\fi}
\reversemarginpar

\begin{document}

\title{Radio sources in the 2dF Galaxy Redshift Survey: AGN, starburst 
galaxies and their cosmic evolution} 
\author{Elaine M. Sadler }
\affil{School of Physics, University of Sydney, NSW\,2006, Australia}
\author{Carole A. Jackson }
\affil{Research School of Astronomy and Astrophysics, The Australian 
National University, Weston Creek, ACT 2611, Australia}
\author{Russell D. Cannon}
\affil{Anglo--Australian Observatory, PO Box 296, Epping, NSW 2121, Australia}

\begin{abstract}
Radio continuum surveys can detect galaxies 
over a very wide range in redshift, making them powerful tools for 
studying the distant universe.  Until recently, though, 
identifying the optical counterparts of faint radio sources and 
measuring their redshifts was a slow and laborious 
process which could only be carried out for relatively small samples 
of galaxies. 

Combining data from all--sky radio continuum surveys with 
optical spectra from the new large--area redshift surveys now 
makes it possible to measure redshifts for tens of thousands 
of radio--emitting galaxies, as well as determining unambiguously 
whether the radio emission in each galaxy arises mainly from an 
active nucleus (AGN) or from processes related to star formation. 
Here we present some results from a recent study of radio--source 
populations in the 2dF Galaxy Redshift Survey, including a new 
derivation of the local star--formation density, and discuss the 
prospects for future studies of galaxy evolution using both radio 
and optical data. 
\end{abstract}

\section{Introduction}
Most astronomical objects evolve on timescales 
which are orders of magnitude longer than a human lifetime, so that 
we effectively view them frozen at a single moment in their life cycle 
and at a fixed orientation.  To piece together the evolutionary 
history of such objects, we therefore need to observe a sample which is 
large enough to cover the full range of luminosity, size, age and 
orientation which exists in nature.  

Advances in technology and data 
processing power mean that surveys of very large numbers of astronomical 
objects are now becoming feasible.  Van den Bergh (2000) recently 
concluded that 
``.. the astronomy of the twenty--first century will be dominated 
by computer--based manipulation of huge homogeneous surveys of various 
types of astronomical objects...''. 

The powerful radio emission from some galaxies acts as a beacon 
which allows them to be seen at very large distances (the median redshift 
of galaxies detected in radio surveys is typically $z\simeq1$; Condon 1989). 
Radio source counts can be used to study the 
cosmological evolution of active galaxies (e.g. Longair 1966, 
Jauncey 1975, Wall et al.\ 1980) but their interpretation is strongly model--dependent. This is especially true at the faint flux densities 
probed by the new generation of all-sky radio imaging surveys 
(NVSS, Condon et al.\ 1998; FIRST, Becker, White \& Helfand 1995; 
WENSS, Rengelink et al.\ 1997; SUMSS, Bock, Large \& Sadler 1999), 
where AGN and `normal' star--forming galaxies both contibute significantly  
(e.g. Condon 1989, 1992).  

The scientific return from 
large radio surveys is enormously increased if the optical counterparts 
of the radio sources can be identified, their optical spectra classified 
(as AGN, starburst galaxy, etc.) and their redshift distribution measured.
Here, we describe some first results from a study of radio--source populations 
in the 2dF Galaxy Redshift Survey.  Much of this work is discussed in more 
detail in Sadler et al.\ (1999, 2002). \\

\begin{figure}[h]
\centerline{\vbox{ 
\psfig{figure=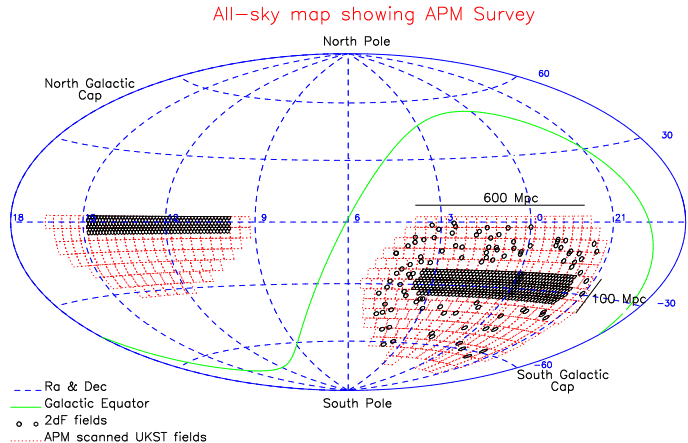,width=9cm,angle=270}
}}
\caption{Sky coverage of the 2dF Galaxy Redshift Survey. The main 
survey regions are the strips around the South Galactic 
Pole near declination $-30^\circ$ and in the northern Galactic 
hemisphere along the celestial equator. } 
\end{figure}

\section{The 2dF Galaxy Redshift Survey } 
The 2dF Galaxy Redshift Survey (2dFGRS) is described by 
Colless et al.\ (2001), and has also been discussed at this meeting 
by Peterson. It used two 
multi--fibre spectrographs at the prime focus of the 4\,m Anglo--Australian 
Telescope (AAT) to measure redshifts for more than 220,000 galaxies in 
1700 deg$^2$ of sky.  The main goal of the 2dFGRS team was to study  
large--scale struture over the redshift range $z=0$ to 0.2, but the high 
quality of the spectra in the public 2dFGRS database means that they 
have also been used for many other studies, such as measuring optical  
luminosity functions for galaxies of different Hubble types (Madgwick 
et al.\ 2002) and examining the effects of environment on the star 
formation rate in galaxies (Lewis et al. 2002).   

\section{Radio imaging surveys in the 2dFGRS area} 
The 1.4 GHz NRAO VLA Sky Survey (NVSS; Condon et al.\ 1998) overlaps 
both the 2dFGRS strips shown in Figure 1, and the southern 
(dec $<-30^\circ$) part of the SGP strip is also covered by the   
843\,MHz Sydney University Molonglo Sky Survey (SUMSS, Bock et al. 
1999). Part of the northern 2dFGRS strip is overlapped by the 
higher--resolution 1.4\,GHz FIRST survey (Becker et al.\ 1995), and 
the properties of the faint 2dFGRS radio sources in this region 
have been discussed by Magliocchetti et al.\ (2002). 

NVSS and SUMSS  have similar resolution ($\sim45$\,arcsec) 
and sensitivity, so in the overlap declination zone $-30^\circ$ to 
$-40^\circ$ they can be combined to measure a radio spectral index 
for sources in common (see Figure 2). The NVSS and SUMSS source 
positions are also accurate enough that we can make unambiguous 
optical identifications down to at least B$\sim20$\,mag.  

\begin{figure}[h]
\centerline{\vbox{ 
\psfig{figure=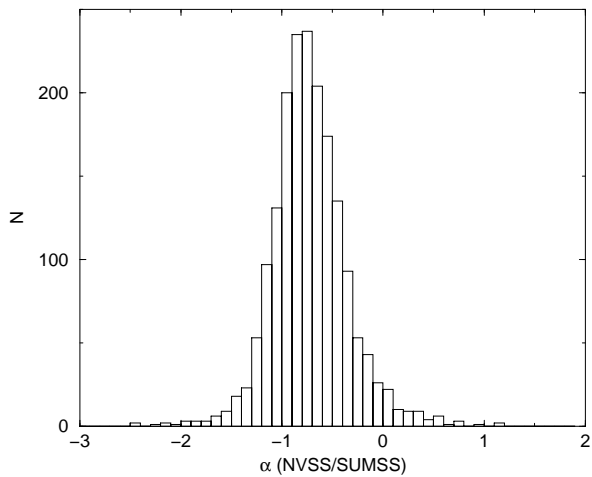,width=7.8cm,angle=0}
}}
\caption{Distribution of radio spectral index $\alpha$ (defined by 
S\,$\,\propto\nu^{\alpha}$ for flux density S at frequency $\nu$, and 
analogous to an optical colour) for radio sources in the overlap zone 
which were detected by both NVSS and SUMSS.  The mean spectral index 
between 843\,MHz and 1.4\,GHz is $-0.734$ (Mauch et al.\ 2002). } 
\end{figure}

The NVSS source density is 60\,deg$^{-2}$ above 2.5\,mJy, and 
the typical 2dFGRS galaxy density is 180\,deg$^{-2}$ above the survey 
cutoff magnitude of $b_{\rm J}=19.4$.  The intersection of the two 
surveys is small, with less than 1.5\% of the 2dFGRS galaxies 
detected as radio sources by NVSS (see Table 1), but the size 
of the 2dFGRS means that this is still the largest and most homogeneous 
set of optical spectra of radio galaxies so far observed. 

\begin{figure}[t]
\centerline{\vbox{ 
\psfig{figure=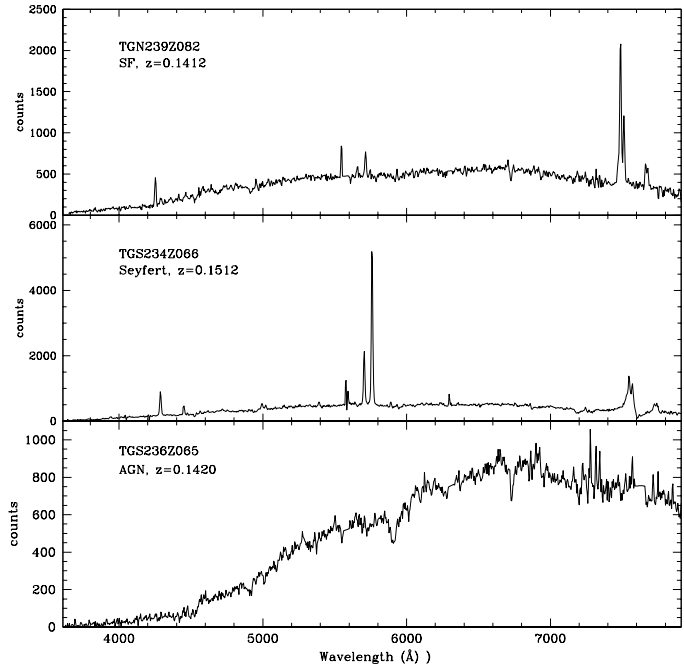,width=9.8cm,angle=0}
}}
\caption{Spectra of three 2dFGRS radio sources with redshift $z\sim0.15$ 
(Sadler et al.\ 1999): 
({\it top}\,) a star--forming galaxy, with strong, narrow Balmer emission 
lines of H$\alpha$ and H$\beta$ and weaker emission lines of 
[SII] 6716,6731, [NII] 6583, [OIII] 4959,5007 and [OII] 3727; 
({\it middle}\,) an emission--line AGN, with broad H$\alpha$ emission 
and [OIII] $>>$H$\beta$; ({\it bottom}\,) a radio galaxy (AGN) with an absorption--line spectrum.} 
\end{figure}
\section{Radio source populations in the 2dFGRS } 
2dFGRS spectra of each of the radio--source IDs were classified visually 
as either `AGN' or `star--forming' (SF). AGN galaxies have either (i) 
an absorption--line spectrum like that of a giant elliptical galaxy, 
(ii) an absorption--line spectrum with weak LINER--like emission lines, 
or (iii) a stellar continuum dominated by nebular emission lines such as 
[OII] and [OIII] which are strong compared with any Balmer--line emission.  
In SF galaxies, strong, narrow emission lines of H$\alpha$ and (usually) H$\beta$ dominate the spectrum.  

As can be seen from Table 1, the 2dFGRS radio--source population is 
a roughly equal mixture of star--forming and active galaxies.  The two 
classes overlap in radio luminosity (though the active galaxies are
on average both more distant and more luminous), and in many cases 
can only be distinguished by examining their optical spectra.  Because 
both populations are well--represented in the 2dFGRS/NVSS data set, 
we can measure accurate radio luminosity functions for both AGN and 
star--forming galaxies as a local benchmark for future 
studies at higher redshift. 

\vspace*{-0.2cm}
\begin{table}
\caption{2dFGRS radio--source statistics }
\begin{tabular}{rl}
& \\
\tableline
\multicolumn{2}{l}{\sl Analysed so far:} \\
58,454 & 2dFGRS spectra (25\% of total) \\
 757 & matched with NVSS radio sources (1.3\%) \\
 441 & radio galaxies (=AGN) \\
 272 & star--forming galaxies (=IRAS galaxies) \\
%  12 & of these are ULIRGs (L$_{\rm FIR} > 10^{12}$\,M$_\odot$) \\
  44 & unclassified spectra (low S/N) \\
\tableline
\tableline
\end{tabular} 
\end{table}

\subsection{Star--forming galaxies}
The 1.4\,GHz radio emission from star--forming galaxies has both a 
thermal component from individual HII regions and a (dominant) large--scale 
non--thermal component arising from synchrotron emission in the disk 
(Condon 1992). Figure 4 shows two star--forming 2dFGRS galaxies which 
were also detected as NVSS radio sources.  

\begin{figure}[h]
\centerline{\psfig{figure=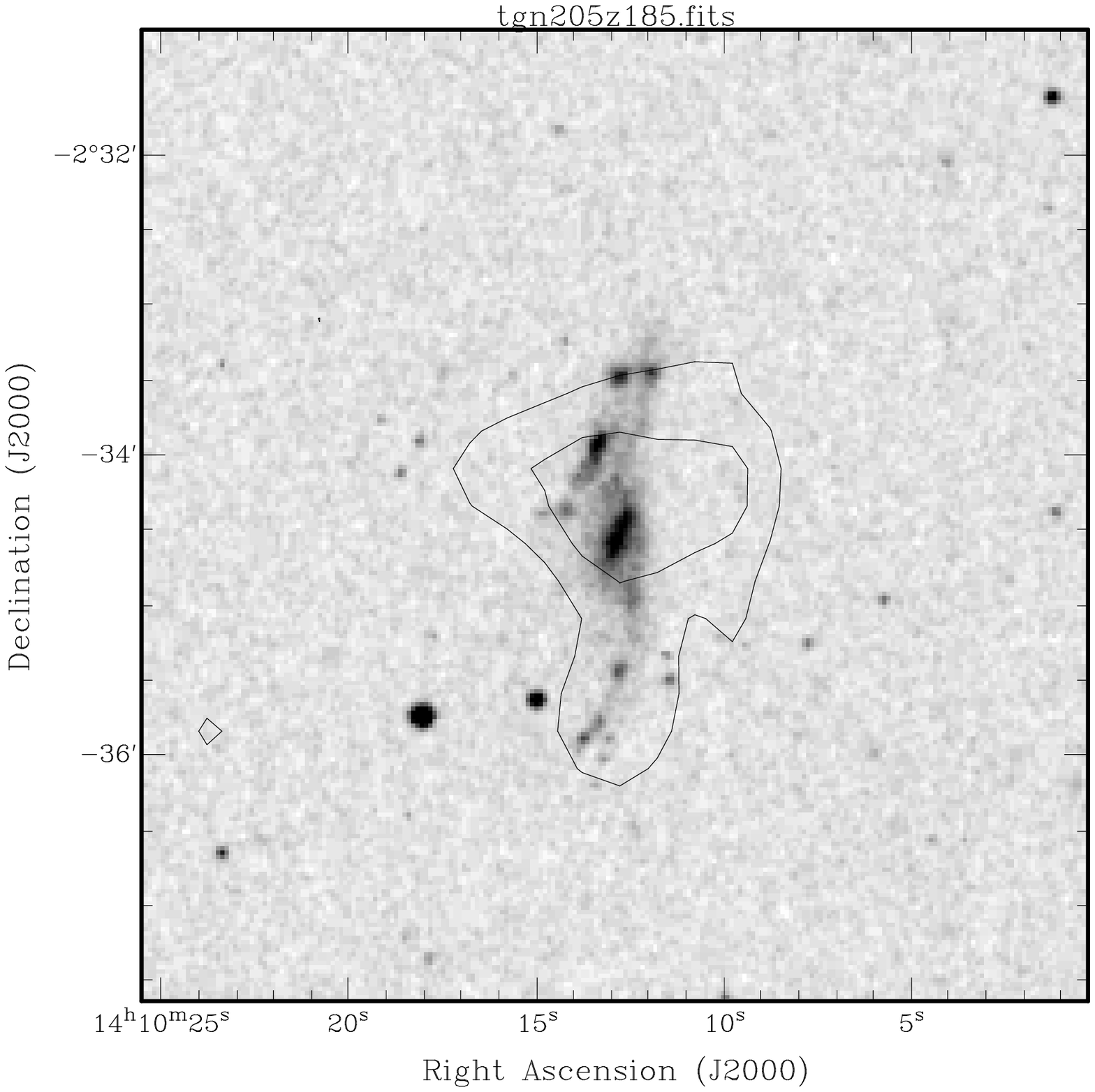,width=4.8cm,angle=0}
\psfig{file=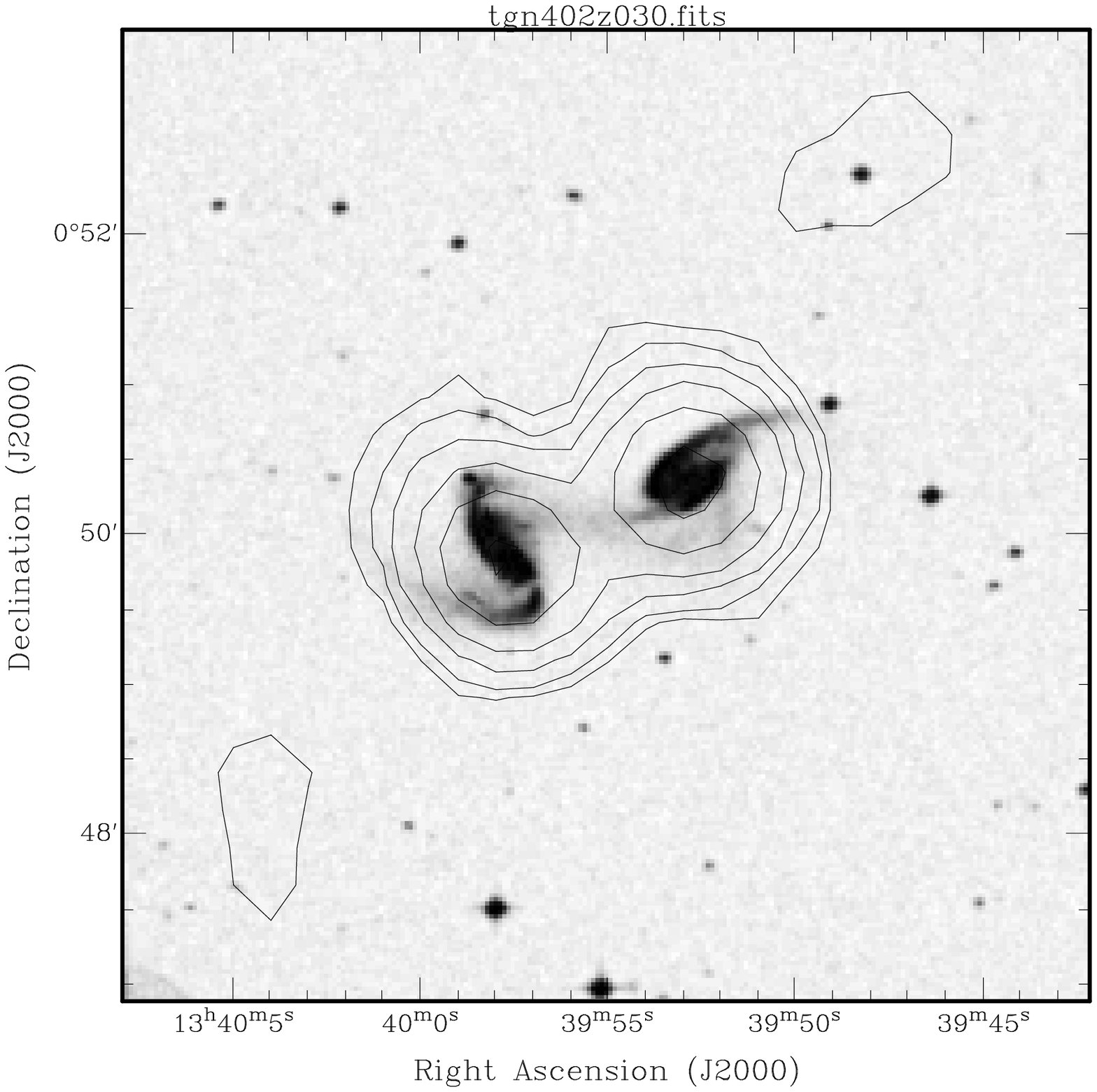,width=4.8cm,angle=0}}
\caption{Contours of 1.4\,GHz radio flux density overlaid on the optical 
images of two star--forming 2dFGRS galaxies.  The estimated star--formation 
rates are ({\it left}\,) 2\,M$_\odot$\,yr$^{-1}$ and ({\it right}\,) 120\,M$_\odot$\,yr$^{-1}$. } 
\end{figure}

Most of the star--forming 2dFGRS galaxies are also detected as far--infrared 
(FIR) sources in the IRAS Faint Source Catalogue (FSC: Moshir et al.\ 1990)  
and follow the well--known FIR--radio correlation (e.g. de Jong et al.\ 1985;  
Helou, Soifer \& Rowan--Robinson 1985). The radio luminosity of these 
galaxies gives an independent estimate of their current star--formation 
rate (e.g. Sullivan et al.\ 2001), and by integrating over the 
radio luminosity function for 2dFGRS/NVSS galaxies we can estimate 
the star--formation density in the local universe in a way which is 
unaffected by dust. 

\begin{figure}[t]
\centerline{\psfig{figure=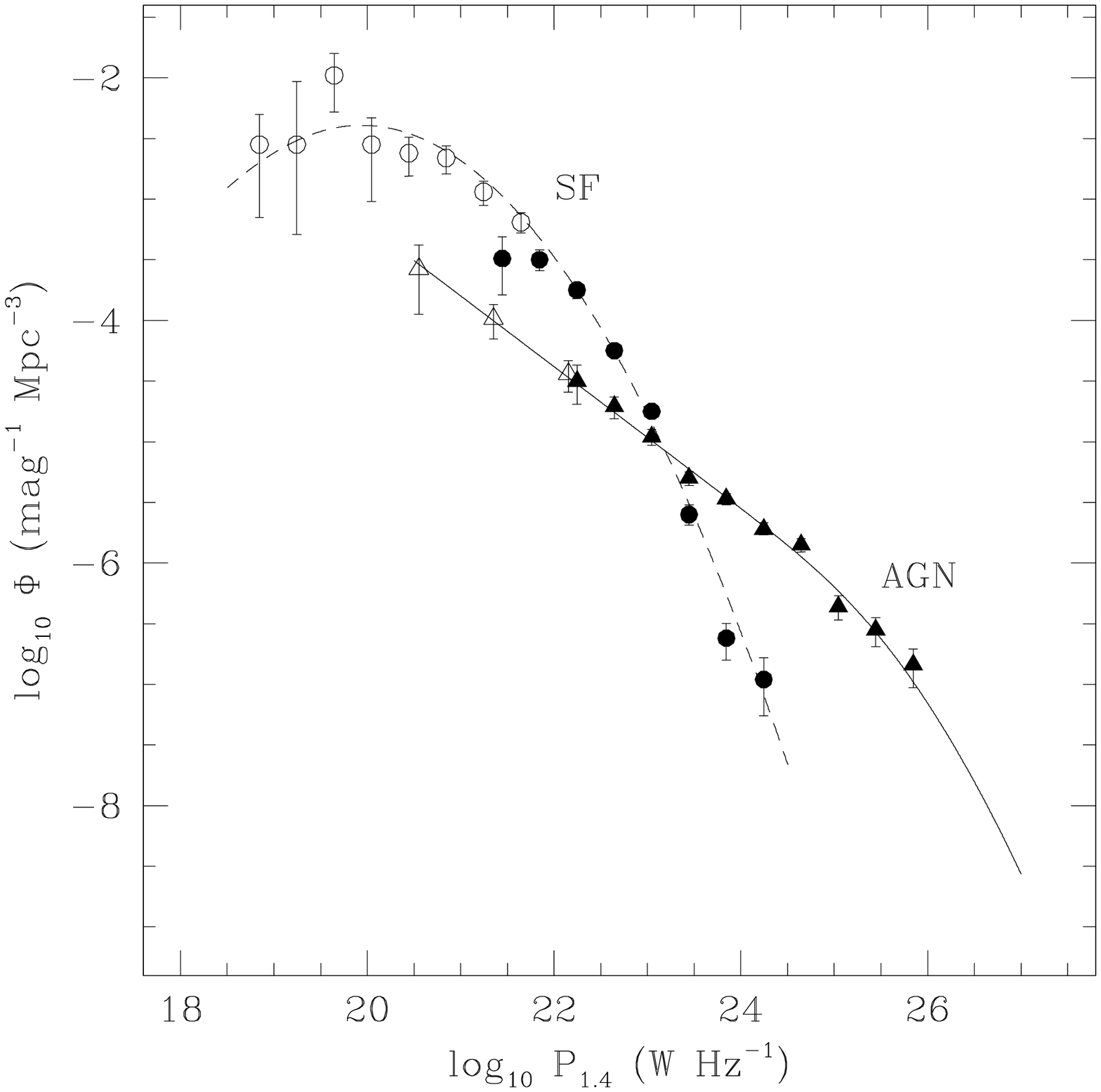,width=6.4cm,angle=0}
\psfig{file=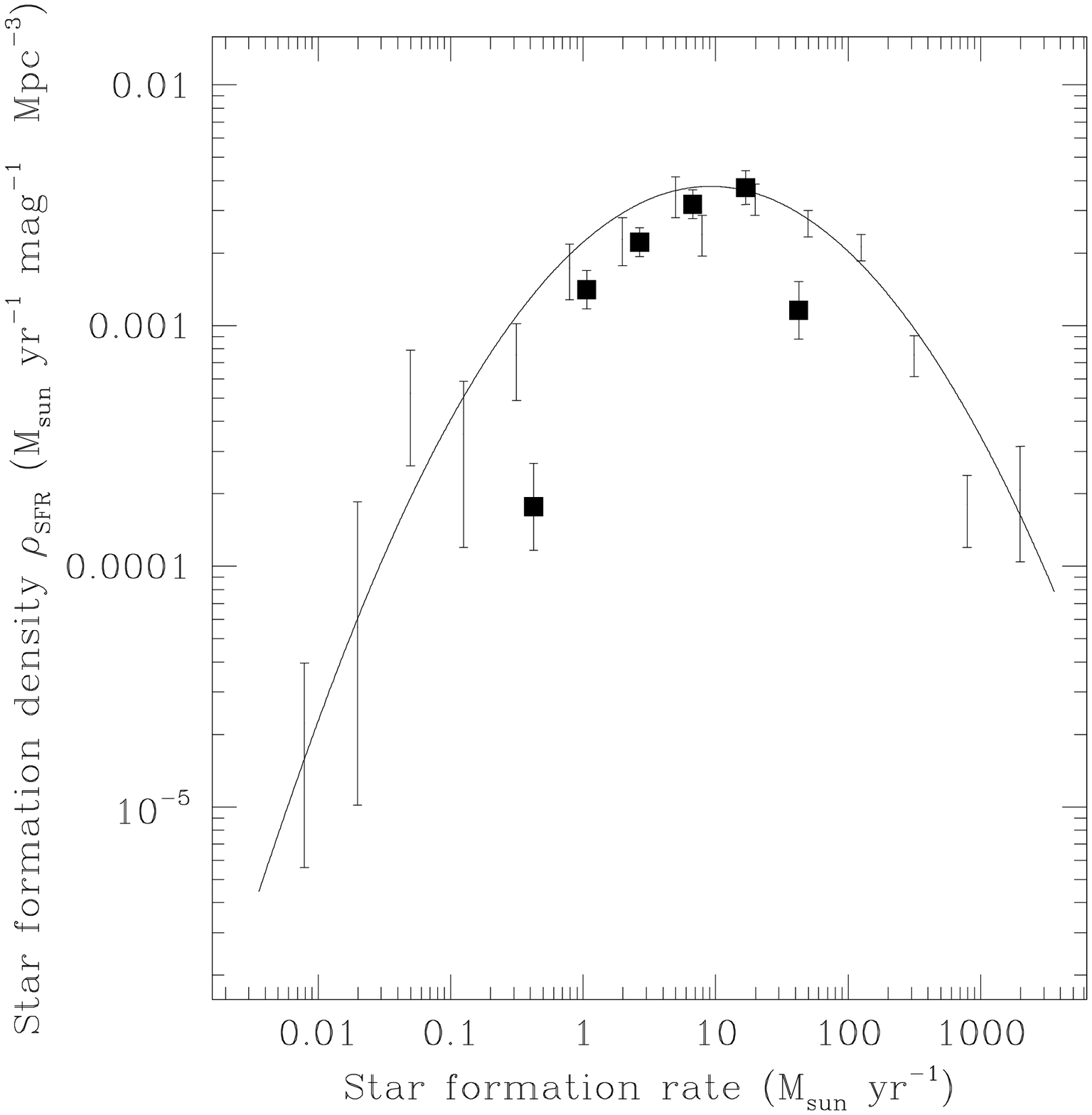,width=6.4cm,angle=0}}
\caption{({\it Left}\,) Local radio luminosity functions 
(H$_0$=50 km\,s$^{-1}$ Mpc$^{-3}$) for star--forming 
(SF) galaxies and AGN, from Sadler et al.\ (2002).  Below 
10$^{22}$\,W Hz$^{-1}$ additional points from nearby bright spiral 
(Condon 1989) and elliptical (Sadler, Jenkins \& Kotanyi 1989) 
galaxies have been added to the 2dFGRS data.  
({\it Right}\,) Local star--formation density 
(in M$_\odot$\,yr$^{-1}$\,mag$^{-1}$\,Mpc$^{-3}$) for galaxies with 
star--formation rates between 0.01 and 1,000\,M$_\odot$\,yr$^{-1}$.
The solid line is derived from the local radio  
luminosity function, and the filled squares are values derived by Gallego 
et al.\ (1995) from H$\alpha$ measurements. } 
\end{figure}

The integrated star--formation density of 
$0.022\pm0.004$\,M$_\odot$\,yr$^{-1}$\,Mpc$^{-3}$ derived from 
the 2dFGRS radio data (Sadler et al.\ 2002) is slightly higher than 
the value of $0.013\pm0.006$ derived optically by Gallego et al.\ (1995). 
The main reason for the difference is that 
the 2dF sample has an excess of galaxies with star formation rates above 
50\,M$_\odot$\,yr$^{-1}$ (see Figure 5).  Condon, Cotton \& Broderick 
(2002) suggest that this difference is due to evolution of the galaxies 
with the highest star--formation rates within 
the 2dFGRS sample volume.  They also note 
that the data are consistent with luminosity evolution of the form 
$f(z) \sim (1+z)^3$ for star--forming galaxies. 

\subsection{Active galaxies}
The radio luminosity function of 2dFGRS active galaxies (see Figure 5) 
has a power--law spectrum of the form 
$\Phi({\rm P_{1.4}}) \propto {\rm P_{1.4}}^{-0.62\pm0.03}$ over almost 
five decades in radio luminosity from 10$^{20.5}$ to 10$^{25}$ W\,Hz$^{-1}$.  

Franceschini, Vercellone \& Fabian (1998) examined the relation between 
galaxy luminosity, black hole mass and radio power in nearby active 
galaxies, and concluded that the radio power of an AGN is both a good 
indicator of the presence of a supermassive black hole and an estimator 
of its mass; 
though Laor (2000) noted that the correlation between radio power 
and black hole mass has a large scatter.  Using a larger data set, 
Snellen et al.\ (2002) confirm the relation found by Francescini et al.\ 
(1998) but note that it applies only for ``relatively passive'' elliptical 
galaxies (i.e. those with radio powers below $\sim10^{24}$\,W\,Hz$^{-1}$).  

If we apply the relation found by Francescini 
et al.\ to the 2dFGRS active galaxies, we derive a value of 
$\rho_{\rm BH}\ =\ 1.6^{+0.4}_{-0.6}\times10^5\ {\rm M}_\odot\ {\rm Mpc}^{-3}$ 
for the total mass density of supermassive 
(M$_{\rm BH}>8\times10^7$\,M$_\odot$) 
black holes in the local universe.  This is close to the value 
of $1.4-2.2\times 10^5$ derived for QSOs by Chokshi \& Turner (1992), 
suggesting that local low--power radio galaxies may be the remnants 
of most or all of the high--redshift QSOs. 

\vspace*{-0.2cm}
\section{PKS\,0019--338, a post--starburst radio galaxy} 
Among the 2dFGRS radio galaxies, we recently discovered the remarkable
object PKS\,0019--338 at $z=0.129$.  Its optical spectrum (Figure 6) 
shows both strong emission lines characteristic of an AGN and an 
underlying blue stellar continuum with strong hydrogen Balmer--line 
absorption. 
The strength of the H$\delta$ absorption line implies that this 
galaxy had a massive starburst which finished only $\sim1.5\times10^8$ 
years ago (K.\ Bekki \& W.J.\ Couch, private communication). 

\begin{figure}[h]
\centerline{\vbox{ 
\psfig{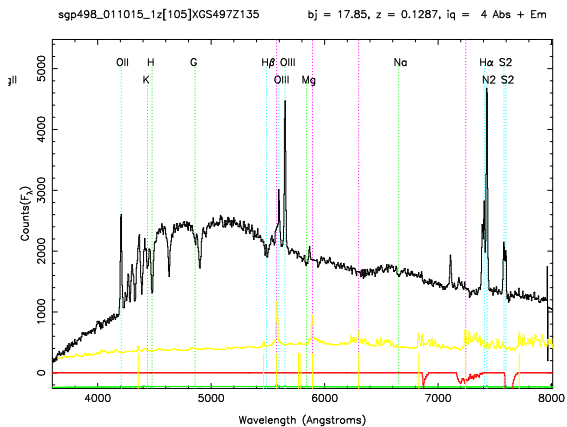}
}}
\caption{2dFGRS spectrum of the `post--starburst' radio galaxy 
PKS\,0019--338. } 
\end{figure}

PKS\,0019--338 belongs to the class of `e$+$A' galaxies, with 
an old stellar population overlaid with a substantial 
younger population created in a massive starburst much less than 1\,Gyr 
ago (e.g. Dressler \& Gunn 1983; Zabludoff et al.\ 1996).  The radio 
source lies within the optical galaxy and has a steep ($\alpha=-1.1$) 
sepctrum, placing it in the class of compact 
steep--spectrum (CSS) radio galaxies.  This implies an age of less than 
10$^6$ years for the radio source (e.g. de Silva et al.\ 1999).  
If the starburst and AGN in this galaxy 
are causally connected, there is a large time 
lag ($\sim10^8$\ years) between the peak of the starburst and the onset 
of radio emission.  

\vspace*{-0.2cm}
\section{Future work } 
We have so far analysed only 25\% of the 2dFGRS radio sample.  
Adding the rest will give us a large 
enough sample to measure the evolution of both AGN and star--forming 
galaxies out to $z\sim0.3$. The 6dF Galaxy Survey described at this meeting 
by Wakamatsu (see also Watson et al.\ 2001) will yield roughly 12,000 
more radio--source spectra to $z\sim0.1$, allowing a detailed investigation 
of the faint end of the radio luminosity function, and will 
also have high enough resolution to measure stellar velocity dispersions. 
By using the velocity dispersion as an independent measure of the mass of 
the central black hole, we should be able to investigate in more detail the 
correlation between black hole mass and radio power in nearby elliptical 
galaxies. 

\acknowledgments
EMS thanks the meeting organisers for financial support, and 
Dr Kenji Bekki and Prof. Warrick Couch for helpful discussions 
on the ages of `post--starburst' galaxies.

\end{document}